\documentclass[CEJP,DVI]{cej} 
\usepackage{layout}
\usepackage{amsmath}
\usepackage{textcomp}
\usepackage{hyperref}

\newcommand{\sqrtsNN}{\mbox{$\sqrt{\mathrm{s}_{_{\mathrm{NN}}}}$}}
\newcommand{\axi}{$\overline{\Xi}^+$}
\newcommand{\xim}{$\Xi^-$}
\newcommand{\alam}{$\overline{\Lambda}$}
\newcommand{\lam}{$\Lambda$}
\newcommand{\ks}{$\mathrm{K}^{0}_{\mathrm S}$}

\newcommand{\ppt}{$p_{\rm T}$}

\usepackage{epsfig}
\usepackage{amssymb} 
\usepackage{lineno} 

\title{\ks, \lam\ and $\Xi$ production at intermediate to high \ppt\ from Au+Au collisions at \sqrtsNN\ =~39, 11.5 and 7.7 GeV}

\articletype{Editorial}

\author{Xianglei Zhu\email{zhux@tsinghua.edu.cn} (for the STAR Collaboration)}

\institute{Department of Engineering Physics, Tsinghua University, Beijing 100084, China}

\abstract{We report on the \ppt\ dependence of nuclear modification factors ($R_{CP}$) for \ks, \lam, $\Xi$ and the \alam/\ks\  ratios at mid-rapidity from Au+Au collisions at \sqrtsNN = 39, 11.5 and 7.7 GeV. At \sqrtsNN = 39 GeV, the $R_{CP}$ data shows a baryon/meson separation at intermediate \ppt\ and a suppression for \ks\ for \ppt\ up to 4.5 GeV/$c$; the \alam/\ks\ shows baryon enhancement in the most central collisions. However, at \sqrtsNN = 11.5 and 7.7 GeV, $R_{CP}$ shows much less baryon/meson separation and \alam/\ks\ shows almost no baryon enhancement. These observations indicate that the matter created in Au+Au collisions at \sqrtsNN = 11.5 or 7.7 GeV might be distinct from that created at \sqrtsNN\ = 39 GeV. 
}

\keywords{strangeness production \*\ heavy ion collisions \*\ QCD phase transition\*\ onset of deconfinement}
\pacs{25.75.Dw,25.75.Nq}

\begin{document}
\maketitle


\section{Introduction}

Hadron production at intermediate to high \ppt\ is a good probe to study medium properties and hadron formation mechanism in heavy ion collisions. At high \ppt, the hard processes, which can be calculated with perturbative QCD, are expected to be the dominate mechanism for hadron productions. Therefore, the nuclear modification factor, $R_{CP}$, which is the ratio of $N_\mathrm{bin}$ scaled particle yields in central collisions relative to peripheral collisions, would equal unity if hard processes were not affected by the presence of the medium. It was observed at RHIC that, at high \ppt, the $R_{CP}$ of various particles is much less than unity \cite{lamont}, indicating the dramatic energy loss of the scattered partons in the dense matter (``jet quenching"). 
$R_{CP}$ of hadrons have been measured at SPS \cite{rcpna49,rcpwa98} as well, though the limited statistics restricts the measurement at relatively lower \ppt( $\lesssim 3$ GeV/$c$). 
By measuring the nuclear modification factor in heavy ion collisions at this energy range, one can potentially pin down the beam energy at which interactions with the medium begin to affect hard partons \cite{Aggarwal:2010cw}. 

At intermediate \ppt, particle production may not be purely from modified jet fragmentation. This was first indicated by the measurement of baryon to meson ratios. It was found at RHIC that, the $p/\pi$ and $\Lambda/$\ks\ ratios are larger than unity for intermediate \ppt\ in more central events, and much higher than that observed in elementary collisions \cite{Adcox:2003nr,lamont,starb2m}. At this intermediate \ppt\ range, the $v_2$ and $R_{CP}$ of baryons are also larger than that of mesons. All of these phenomena hint toward different hadronization mechanisms at this \ppt\ range. There are recombination/coalescence models which allow soft partons to coalesce into hadrons, or soft and hard parton to recombine into hadrons \cite{coal}. They naturally repoduce an enhanced baryon to meson ratio when the parton \ppt\ distribution is exponential. Such models require constituent quarks to coalesce or recombine, and hence the observation of coalescence or recombination behavior is one of the corner-stone pieces of evidence for the formation of the sQGP. It is also interesting to investigate in which collision energies these phenomena are prevalent \cite{Aggarwal:2010cw}, in order to locate the energy point at which the onset of the deconfinement happens.

STAR has collected high statistics Au+Au data at \sqrtsNN\,\,=~39, 11.5 and 7.7~GeV during the first phase of the Beam Energy Scan (BES) program in 2010. This allows high precision strangeness measurements, especially at intermediate to high \ppt, at these energies. In STAR, strange hadrons are reconstructed with their secondary TPC tracks through the topology of their weak decay channels, \ks$\,\,\rightarrow \pi^+\pi^-$, $\Lambda\rightarrow p\pi$ and charged $\Xi\rightarrow\Lambda\pi$. In this paper, we will focus on the nuclear modification factors ($R_{CP}$) of these strange hadrons and \alam/\ks\ ratios at three BES energies. This analysis is based on about 14, 12 and 5 million minimum bias Au+Au events at \sqrtsNN\ = 39, 11.5 and 7.7 GeV, respectively. 

\section{Results}

\begin{figure} [h]
\centering
\includegraphics[width=1.0\textwidth]{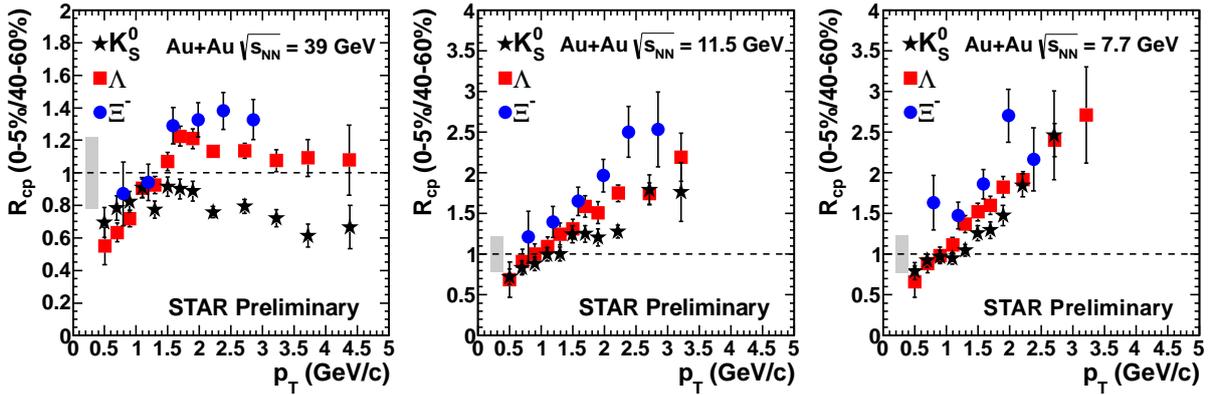}
\caption{The nucler modification factor, $R_{CP}$, as a function of \ppt\ for \ks, \lam\ and \xim\ at $|y|<0.5$ from Au+Au collisions at \sqrtsNN\ = 39, 11.5 and 7.7 GeV. Errors are statistical only. } \label{fig1}
\end{figure}

After correcting the raw spectra for geometrical acceptance and reconstruction efficiencies, we obtain the \ppt\ spectra of \ks, \lam(\alam), \xim(\axi)  at mid-rapidity ($|y|<0.5$) in different centralities for Au+Au collisions at \sqrtsNN\ = 39, 11.5 and 7.7 GeV \cite{zhusqm11}. The \lam(\alam) spectra have been corrected for the feed-down contributions from $\Xi$ and $\Xi^0$ weak decays. The percentage of the feed-down contribution in the raw \lam(\alam) spectra depends on the \lam(\alam) topological selection cuts, and decreases with the increase of \ppt\ and the decrease of centrality. With the current cuts, the overall feed-down contributions for minimum bias events (0-80\%) are 20\%~(25\%), 15\%~(27\%) and 11\%~(35\%) for \lam~(\alam) at \sqrtsNN\,=\,\,39, 11.5 and 7.7 GeV, respectively.  

Figure \ref{fig1} shows the $R_{CP}$ as a function of \ppt\ for \ks, \lam\ and \xim\ 
at $|y|<0.5$ from Au+Au collisions at \sqrtsNN = 39, 11.5 and 7.7 GeV. 
The shaded area shows the systematic error of $R_{CP}$ due to the uncertainty of $N_\mathrm{bin}$ estimation with a Monte Carlo Glauber model. Figure \ref{fig1} (left) shows that at  \sqrtsNN = 39 GeV, the $R_{CP}$ of \ks\ increases and reaches a maximum value of 0.9 at \ppt\ $\sim$ 1 GeV/$c$, then decreases steadily to around 0.6 at the maximum achievable \ppt\ of 4.5 GeV/$c$. For \lam, the $R_{CP}$ reaches a maximum value of 1.2 at \ppt\ around 2 GeV/$c$, then decreases slowly to about 1 at \ppt\ $\sim$ 4.5 GeV/$c$. The \xim\ $R_{CP}$ are similar to that of \lam, but lacks statistics for \ppt\ above 3 GeV/$c$.  Figure \ref{fig1} (left) shows a suppression for \ks\ up to 4.5 GeV/c, and baryon's $R_{CP}$ much larger than meson's at intermediate \ppt. This observation of baryon/meson separation is consistent with that from heavy ion collisions at higher RHIC energies \cite{lamont}. The middle and right plots in Fig. \ref{fig1} show the corresonding $R_{CP}$ data from Au+Au collisions at \sqrtsNN\ = 11.5 and 7.7 GeV. Though the maximum accessible \ppt\ is smaller at these two energies, the $R_{CP}$ data seems qualitatively different from that at \sqrtsNN\ = 39 GeV: there is no suppression for \ks\ at \ppt\ above 2 GeV/$c$; and there is no clear baryon/meson separation at intermediate \ppt. The maximum $R_{CP}$ of three hadrons are all much larger than unity, which indicates that the initial state effects, such as the Cronin effect \cite{cronin}, might be dominant at these two energies.

\begin{figure} [h]
\centering
\includegraphics[width=1.0\textwidth]{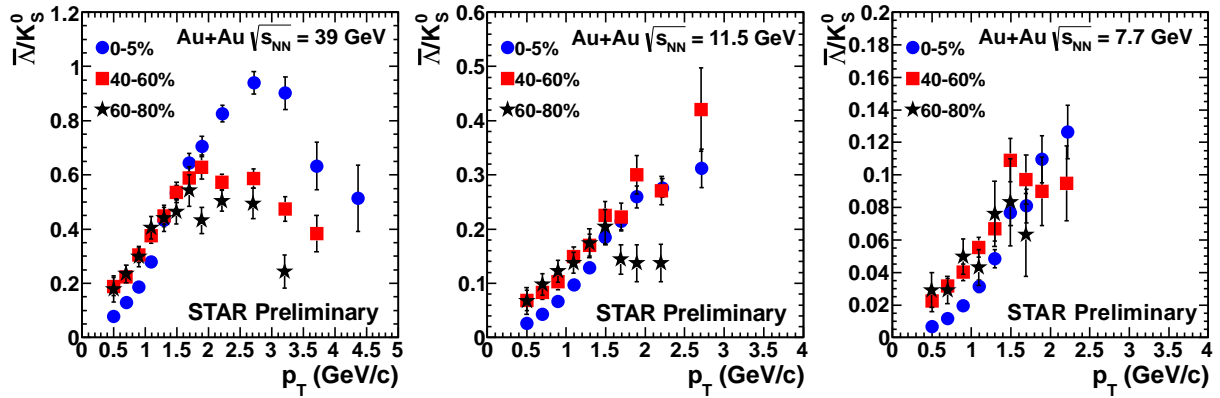}
\caption{The baryon to meson ratio, \alam/\ks, as a function of \ppt\ for 0-5\%, 40-60\% and 60-80\% centralities 
at $|y|<0.5$ from Au+Au collisions at \sqrtsNN\ = 39, 11.5 and 7.7 GeV. Errors are statistical only.  }
\label{fig2}
\end{figure}

Figure \ref{fig2} shows the \alam/\ks\ ratios as a function of \ppt\ in three centralities from Au+Au collisions at \sqrtsNN
= 39, 11.5 and 7.7 GeV. For heavy ion collisions at these energies, it was suggested to look at the \alam/\ks\ ratios instead of \lam/\ks\ to make a comparison on baryon enhancement to higher energies \cite{alice}, due to the large net-baryon density at mid-rapidity at these lower energies. As shown in Fig. \ref{fig2} (left), at \sqrtsNN=39 GeV, the \alam/\ks\ can reach a maximum value of unity at \ppt\ $\sim$ 2.5 GeV/$c$ in the most central collisions, while in the peripheral collisions, the maximum value is only about 0.5. This shows that the baryon enhancement at intermediate \ppt\ at 39 GeV is similar to that observed at higher energies. However, as shown in the middle and right plots of Fig. \ref{fig2}, in Au+Au collisions at \sqrtsNN\ = 11.5 and 7.7 GeV, the maximum values of \alam/\ks\ at the measured \ppt\ range is much smaller than unity, and the difference between central and peripheral collisions is also less significant. This shows almost no baryon enhancement at intermediate \ppt\ in Au+Au collisions at \sqrtsNN\ = 11.5 and 7.7 GeV.  

\section{Summary}

In summary, we presented the \ppt\ dependence of the nuclear modification factor ($R_{CP}$) for \ks, \lam, $\Xi$ and the \alam/\ks\ ratios at mid-rapidity from Au+Au collisons at \sqrtsNN = 39, 11.5 and 7.7 GeV. At \sqrtsNN\ = 39 GeV, the $R_{CP}$ data shows a baryon/meson separation at intermediate \ppt\ and suppression for \ks\ for \ppt\ up to 4.5 GeV/$c$; the \alam/\ks\ shows baryon enhancement in the most central collisions. However, at \sqrtsNN\ = 11.5 and 7.7 GeV, $R_{CP}$ shows much less baryon/meson separation and \alam/\ks\ shows almost no baryon enhancement at intermediate \ppt. These observations indicate that the matter created in Au+Au collisions at \sqrtsNN\ = 11.5 and 7.7 GeV might be distinct from that created in \sqrtsNN\ = 39 GeV. STAR has accumulated high statisticis Au+Au data at \sqrtsNN\ = 19.6 and 27 GeV in 2011. The strangeness production at these two energies will be helpful to further pin down the transition energy point.
\\
{\bf Acknowledgments:} X. Zhu thanks the support by NSFC (Grant Nos. 10905029, 11035009) and the Foundation for the Authors of National Excellent Doctoral Dissertation of P.R. China (FANEDD) (No. 201021).


\end{document}